\documentstyle[floats,aps,epsf,prl]{revtex}

\def \be{\begin{displaymath}}
\def \ee{\end{displaymath}}              

\def \ben{ \begin{equation} }
\def \een{ \end{equation}   }            

\def \bea{\begin{eqnarray*}}             
\def \eea{\end{eqnarray*}}

\def \bean{\begin{eqnarray}}             
\def \eean{\end{eqnarray}}

\def \nn{\nonumber}

\def \Re{ \mbox{Re} }

\def \N#1#2{ \mbox{N}_{#1,#2} }
\def \Nprime#1#2{ \mbox{N}'_{#1,#2} }

\def \Ref#1{(\ref{#1})}

\def \eps{\varepsilon}

\def \vrho{\varrho}
\def \tphi{\tilde \phi}
\def \lambar{\bar \lambda}

\def \e{ {\rm e}}

\def \inv{ ^{-1} }

\def \invb#1 { \frac{1}{#1} }

\def \av#1{ {\left\langle #1 \right\rangle} }

\def \dx{\partial_x}

\def \fr#1#2{ \frac{#1}{#2} }

\begin{document}
\draft 
\twocolumn[\hsize\textwidth\columnwidth\hsize\csname
@twocolumnfalse\endcsname

\title{Coulomb drag between quantum wires with different electron
densities}

\author{Thomas Fuchs and  Rochus Klesse}

\address{ Universit\"at zu K\"oln, Institut f\"ur Theoretische Physik,
     Z\"ulpicher Str. 77, D-50937 K\"oln, Germany}

\author{Ady Stern}

\address{ Department of Condensed Matter Physics, The Weizmann
Institute of Science,
Rehovot 76100, Israel}

\date{August 5, 2004}

\maketitle

\begin{abstract}
  We study the way back-scattering electron--electron interaction
generates Coulomb drag between quantum wires with different densities. 
At low temperature $T$ the system can undergo a
commensurate-incommensurate transition as the potential difference
$|W|$ between the two wires  
passes a critical value $\Delta$, and this transition is reflected in
a marked change in the dependence of drag resistivity on $W$ and
$T$. At high temperature a density difference between the wires
suppresses Coulomb drag induced by back scattering, and we use the
Tomonaga--Luttinger model to study this suppression in detail.   
\end{abstract}

\pacs{PACS}]

\section{Introduction}\label{sec-introduction}

A constant electric current in a conductor establishes
a stationary non-equilibrium distribution of electron momenta.
Relaxation processes with electrons in a second (``passive'') conductor
nearby
can result in a significant drag force, which leads to a measurable
drag current
or drag voltage in the passive conductor. This effect is known as
Coulomb drag if
the momentum transfer is mediated by the Coulomb interaction
\cite{coulomb_drag}.
The drag effect is very sensitive to electronic correlations and can
therefore be used to probe them.

While the Coulomb drag between two-dimensional electron systems
has been extensively studied experimentally \cite{rojo}, the
observation of this
effect in strictly one-dimensional (1D) systems still presents a
challenge and
experimental work on this subject is quite limited
\cite{debray,debray_review,yamamoto}. The 1D case is
especially interesting since here electronic correlations are much
stronger than in two- or three-dimensional systems. This is reflected
by the fact that
the Fermi-liquid theory, which applies in three dimensions and
marginally
holds for two dimensional systems, generally fails in one dimension.
Instead,
essential features of a correlated 1D electronic system are captured by
the
Tomonaga-Luttinger (TL) model \cite{voit}.

A theory for the 1D Coulomb drag that is based on the TL model
\cite{nazarov_averin,klesse_stern} predicts a behaviour that
qualitatively deviates from the one in higher dimensions:
At temperatures $T$ below a certain
energy gap $\Delta$ the drag
resistivity $\rho_D$ between long wires of equal electron density
increases exponentially with decreasing temperature,
$\rho_D \propto \exp \Delta / T$ \cite{units}.
At temperatures well above the gap the drag
resistivity shows a
distinct power law behaviour, $\rho_D \propto T^x$, where the
exponent $x$ is determined by the forward scattering
part of intra- and inter-wire interaction.
For vanishing forward scattering interaction $x$ equals unity (and the
linear
$T$-dependence of a Fermi-liquid approach
\cite{hu_flensberg-gurevich_raichev} is reproduced), while it
decreases
with increasing repulsive electron-electron interaction.
The gap energy $\Delta$ corresponds to a correlated ground state,
where the electrons in the two wires order in a zigzag formation with
period equal to the Fermi wavelength $\lambda_F$.
The gap energy $\Delta$ therefore strongly depends on the
backscattering ($2 k_F$) component of the $ee$-interaction.
It is exponentially suppressed if the distance $d$ between the wires
is large compared to $\lambda_F$.

Under the strictly linear spectrum assumed by the TL model a drag
response is generated only by inter-wire back-scattering.
A recent ana\-ly\-sis \cite{pustilnik} by Pustilnik and collaborators
addresses drag induced by forward scattering in a wire with non-linear
dispersion.
In particular, it is shown in that work that at low temperatures
forward
scattering gives a contribution $\rho_D^{(f)}\propto T^2$ to the drag
resistivity if the difference in
the Fermi velocities $\delta v_F$ is small
compared to $T/k_F$. For $\delta v_F$ larger than $T/k_F$,
the contribution due to forward scattering decays more rapidly,
$\rho_D^{(f)} \propto T^5$.

This communication focuses on the backscattering drag between
wires of different electron densities. In principle, the difference in
densities is expected to suppress inter-wire $2k_F$ scattering, since
the momentum $2k_F$ is different in the two layers, and to affect drag
also through the difference in Fermi velocities in the two wires. Here,
we neglect the difference in Fermi velocities between the wires (a
difference which induces corrections of the order of
$\frac{\mu_2-\mu_1}{\mu_1+\mu_2}$, with $\mu_i$ the electro-chemical
potential of the $i$'th wire), and focus on the effect of the difference
between the relevant $2k_F$'s (a difference which induces a much larger
correction of the order of $\frac{\mu_2-\mu_1}{T}$). We envisage the
experimental
situation of two identical wires on  different electro-static
potentials
\cite{debray,yamamoto}.

We start by analyzing Coulomb drag in the high temperature regime,
where inter-wire back-scattering may be treated perturbatively. In
Sec.~\ref{sec-high-temperature} we reinvestigate the drag in
the incommensurate phase with different electron densities
and present detailed results for the non-linear drag at finite
temperatures. Parts of these results are already published in the
review \cite{debray_review}.

Following that analysis, we address the low temperature regime. For
equal densities, the low temperature regime is that of an interlocked
crystal, characterized by a drag resistance that diverges at zero
temperature. As an electro-chemical potential difference $W$ is turned
on between the wires, the relative density between the wires is
initially incompressible, until an  incommensurate-commensurate (IC)
transition \cite{pokrovsky,schulz} takes place \cite{starykh}. This
transition markedly influences the drag.
The transition from the incommensurate to the commensurate
phase is, roughly speaking, a readjustment
of charges into an energetically more favorable zigzag
configuration.

We employ the theory of the IC transition \cite{pokrovsky,schulz} and
use a simple Drude model for solitonic charge transport to investigate
the drag at low temperatures $T \ll \Delta$.
We conclude that for $|W| < \Delta$ the drag resistivity $\rho_D$ is
exponentially large, $\rho_D \propto \exp(\Delta-|W|)/T$.  Above the
transition point the drag shows an inverse square root behaviour,
$\rho_D \propto (W^2 - \Delta^2)^{-1/2}$, which changes to an $1/|W|$
dependence for large $|W| \gg \Delta$ (c.f.\ Fig.\ \ref{fig1}).
This is studied in Sec.~\ref{sec-ic-transition}.

We hope that the present work may help to better understand present
and near-future drag experiments. In particular, together with the
recent results of Pustilnik {\em et al.} \cite{pustilnik} it may help
to clarify the
role of forward and backward scattering in the drag effect.

\section{Coulomb coupled double wire} \label{sec-double-wire}
We consider electrons in two parallel, strictly 1D wires of
length $L$ separated by a distance $d$. We assume the wires to be
identical, and assume that a voltage $W/e$ is imposed between them,
leading to a difference between their electro-chemical potentials
$W=\mu_2-\mu_1$. 
It is assumed that $L$ is much larger than
all other relevant lengthscales of the problem, i.e. the
wires are practically of infinite length.
For the beginning we neglect the electron spin.

Using the same notations as in \cite{klesse_stern} and assuming
that the electro-static potential is constant along each
wire, kinetic and potential energy of the electrons are given by
\be
H_0 =  \sum_{rwk}  v_F (rk - k_F) \: a_{rw}^\dagger (k) a_{rw}(k) +
 \frac{1}{2} W(N_2-N_1)
\ee
The index $r$ corresponds to left ($r=-$) and right ($r=+$) moving
electrons, $w$ refers to active ($w=1$) and passive ($w=2$) wire.
The electron number in wire $w$ is $ N_w = N_{+w} + N_{-w}$, where
$ N_{rw} =  \sum_k a_{rw}^\dagger(k) a_{rw}(k) $.
The 
Fermi wave-number $k_F$ is defined via the identical equilibrium
density
in both wires for vanishing external potentials.

The expressions for forward scattering and backward scattering
electron-electron interactions $H_f$ and $H_b$, resp., are precisely as
in
Ref. \cite{klesse_stern} Sec. IIIB, and are therefore not repeated
here.

We proceed as in \cite{klesse_stern} with a standard transformation
\cite{voit} to bosonic fields
\bea
\phi_w(x)  &=&
 -\fr{i\pi}{ L}  \sum_q \fr{e^{-iqx - \fr{\alpha |q|}{2}}}{q} [
\varrho_{+w}(q) +
 \varrho_{-w}(q)]\\
& &  - \fr{\pi}{ L} N_w x \: ,
\eea
with their conjugated
\bea
\Pi_w(x)  &=&
 \fr{1}{L}  \sum_q e^{-iqx - \fr{ \alpha |q|}{2}}[ \varrho_{+w}(q) -
 \varrho_{-w}(q)] \\
& & + \fr{1}{L} J_w\: ,
\eea
where $J_w = N_{+w} - N_{-w}$, and $\vrho_w$ is the electron density
in
wire $w$.
Note that in the present
case the particle number $N_w$ may change with variation of the
external potentials and therefore the zero-mode terms must be included
here. In the following, the inverse momentum cut-off $\alpha$ is set
to
the Fermi wavelength $2\pi / k_F$.
Then, transformation to  symmetric ($+$) and antisymmetric ($-$)
charge
modes $\phi_{c\pm} = 2^{-1/2} ( \phi_1 \pm \phi_2)$ decouples the
Hamiltonian into independent parts,  $H = H_{c+} + H_{c-}$, with
\bean
H_{c-}  =
  \frac{ u }{2\pi}\int K \pi^{2}\Pi_{c-}^{2} +
  \frac{1}{ K }(\partial_{x}\phi_{c-})^2 dx \nn \\
+ \frac{\bar \lambda E_0}{\pi\alpha} \int  \cos \sqrt{8}\phi_{c-} dx
- \frac{ W}{ 2 \pi} \int \dx \sqrt{8} \phi_{c-}  dx,
\label{grand_canonical}
\eean
and a similar expression without $\cos$-term for $H_{c+}$.
The parameters $u \equiv u_{c-}$ and $K \equiv K_{c-}$ are determined
by the
interwire and intrawire couplings $g_i$ and $\bar g_i$ as explained
in \cite{klesse_stern}. $\bar \lambda = \bar g_1 /2 \pi u$ denotes the
dimensionless interwire backscattering coupling, $E_0 = u/\alpha$ is
of order the Fermi energy.

\section{High temperature drag}
\label{sec-high-temperature}
At high temperatures the two wires are only weakly correlated, and
drag may be calculated perturbatively, in a method similar to that
employed in \cite{nazarov_averin,klesse_stern}. Here we will focus on
the case of wires with different densities.

In the weakly coupled regime 
it is appropriate to switch to a
statistical ensemble with fixed electron numbers  $N_w = L n_w$ ($w=
1,2$).
This enables us to define {\em periodic} bosonic fields by
\be
\tphi_{c-} (x) =  \phi_{c-} (x) + q x\:,
\ee
where
$ q = 2^{-1/2} \pi ( n_1 - n_2) = 2^{-1/2} (k_{F1} - k_{F2}) $.
The dynamics of $\tphi$ follows from Eq. \Ref{grand_canonical} to be
given by the Hamiltonian
\bean
H_{c-}  =
  \frac{ u }{2\pi}\int K \pi^{2}\Pi_{c-}^{2} +
  \frac{1}{ K }(\partial_{x}\tphi_{c-})^2 dx \nn \\
+ \frac{\bar \lambda E_0}{\pi\alpha} \int  \cos \sqrt{8}(\tphi_{c-}-qx)
dx
+ {\rm const.}
\label{micro_canonical}
\eean

We calculate the drag using the formalism of
Refs. \cite{nazarov_averin} and \cite{klesse_stern}, 
whereby the drag voltage is
\ben
\fr{eV_{D}}{L} = - \fr{1}{2 \kappa L} \int_{0}^{L} \langle
  \partial_{x} n_{-}\rangle_{I}   
= \fr{1}{ \sqrt{8} \pi \kappa } 
\langle \partial _{x}^{2}\phi _{c-}\rangle_I \label{draggeneral}\:.
\een
Here, $n_- = (n_1 - n_2)/2$ and $\kappa = K / 2\pi u$ are density and 
compressibility of the anti-symmetric channel, and the averaging is
taken
with respect to a state carrying a current $I$ in the active wire (1).

Using the condition of stationarity and the equation of motion for
$\tphi$ 
we find that to second order in the backward scattering the drag
$\eps_D \equiv eV_D/L$ is 
\begin{eqnarray} \label{dragsinsin}
\eps_D  &=& \fr{4}{\pi} \left (\frac{\bar{\lambda}E_0}{\alpha }\right
)^{2}  \times  \\
& & {\mbox{Im}}\int^{\infty}_{0} \!\! \!\!\!dt \!\!\int \!\! dx 
\langle \cos( \sqrt{8} \tphi_{x,t} - Q x + \omega t) 
\sin( \sqrt{8} \tphi_{0,0} ) \rangle \:, \nn
\end{eqnarray}
where $Q \equiv u \delta k_F$, and the term $\omega t \equiv  -
\fr{\sqrt{2} I}{e} \: t$ results from a Galilei
transformation of the first wire which generates the current $I$
(cf. Refs. \cite{nazarov_averin,klesse_stern}).

Calculating the correlator in the usual way, we obtain
\bean
\eps_D &=& \fr{ E_0^2 }{\pi e u} \lambar^2 \label{drag_field}
\left( \fr{\pi T}{E_0} \right)^{4K-2} \times \\
& & 
\left(
\N{-\fr{{\omega + Q} }{T}}{K} \N{-\fr{ \omega - Q }{T}}{K}
-\N{\fr{ \omega + Q }{T}}{K} \N{\fr{ \omega - Q }{T}}{K}
\right)\:, \nn
\eean
where the function $\N{r}{K}$ is given by
\bean\label{Ndef}
\N{r}{K}  &=&  \lim_{\delta \to 0+} \int ds \: e^{-i s r / \pi}
\left( (\fr{\delta}{s} + i ) \sinh s \right)^{-2K} \\
&=& 2^{2K} \Gamma(1-2K) \Re\left[ \fr{\e^{i\pi K}
    \Gamma(K-i\fr{r}{2\pi})}{\Gamma(1-K-i\fr{r}{2\pi})} \right] \:.
\label{Nexplicit}
\eean
The result \Ref{drag_field} 
is limited to the weakly coupled regime, which
requires $\Delta \ll T$. Further, we have
assumed that all three relevant energies $T$, $|\omega|$, and $|Q|$
are
small compared to $E_0$. Boundary effects are also neglected, which is
allowed as long as the wire length $L$ exceeds the thermal wavelength
$u/T$.

\begin{figure}
  \begin{center}
    \epsfxsize=6.0cm
    \leavevmode
    \epsffile{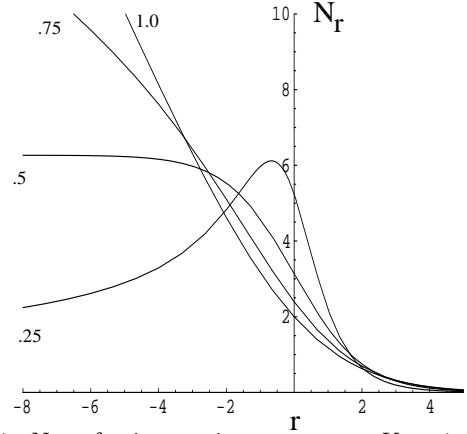}
    \caption{$\N{r}{K}$ for interaction parameter $K=1,\: 0.75,\: 0.5$
and
    $0.25$.
      }
    \label{fig-Nplot}
  \end{center}
  \vspace{-0.5cm}
\end{figure}

As a forward to a discussion of the drag
\Ref{drag_field} as a function of the parameters $T$, $Q$ and
$\omega$, we first analyze the function $\N{r}{K}$.
An asymptotic expansion of Eq. \Ref{Nexplicit} shows that
$\N{r}{K}$ decays exponentially $\sim \e^{-r}$ for large
positive argument $r$, and that  $\N{r}{K} \sim r^{2 K -1}$ for large
negative $r$. More precisely, we find
\be
\N{r}{K} \approx 2 \sin(2 \pi K) \Gamma(1-2K)
\left| \fr{r}{\pi} \right|^{2K-1} \label{asymptotic}
 \times
\left\{
{ 1,\quad  -r \gg 1  \atop e^{-r}, \quad r \gg 1}
\right.
\ee
For the particular values $K=1$, corresponding to non-interacting
electrons, and
$K=1/2$ the expression \Ref{Nexplicit} simplifies to
\bea
\N{r}{1} &=& \fr{2r}{e^r - 1}\:, \\
\N{r}{1/2} &=& \fr{2 \pi}{ e^r + 1}\:.
\eea
While $\N{r}{1}$ bears some resemblance to the
Bose distribution, $\N{r}{1/2}/2\pi $ is the Fermi function
(cf.~Fig.~\ref{fig-Nplot}).

\subsection{Non-linear drag}
\begin{figure}
  \begin{center}
    \epsfxsize=8.0cm
    \leavevmode
    \epsffile{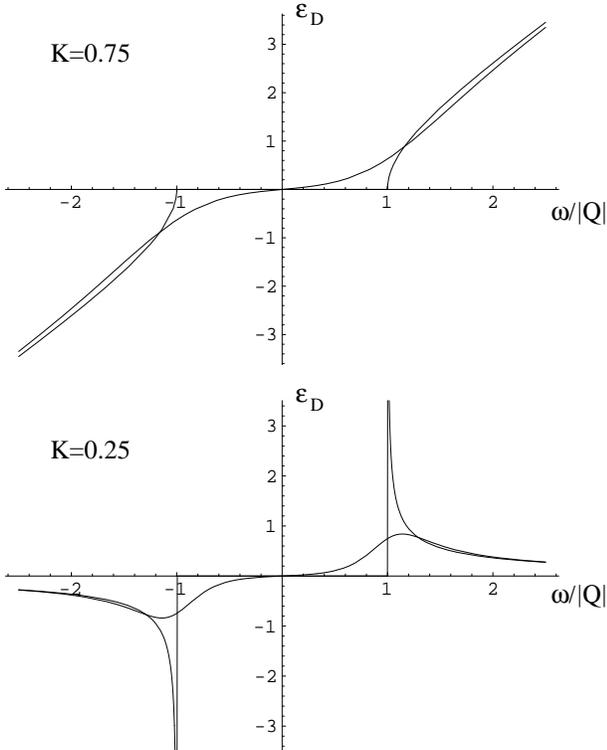}
    \caption{
      Drag in arbitrary units as a function of $\omega \propto I$ in
      units of $|Q| =u |\delta k_F|$ for interaction parameters
      $K=0.75$ and 0.25 at temperatures
      $T=0.25 Q$ (smooth curves) and $T \ll Q$.
      For large currents ($|\omega| \gg |Q|$) the drag goes as
      $\omega^{4K-2}$. At low temperatures the drag shows an
      $|\omega-Q|^{2K-1}\Theta(\omega^2 - Q^2)$ singularity near the
threshold
      $|\omega| \sim |Q|$, which is smeared over a regime where $|
      |\omega| - |Q|| \sim T$.
      }
    \label{fig-drag}
  \end{center}
  \vspace{-0.5cm}
\end{figure}

For $T \ll |\omega|$ and $T\ll |Q|$ (but still $T \gg \Delta$ and $T\gg
u/L$ )
we can approximate $\N{r}{K}$ in Eq. \Ref{drag_field} by the asymptotic
expression
for negative $r$, and $\N{r}{K}=0$ for positive argument
$r$.
This shows that for $|\omega| - |Q| \gg T $
\bea
\eps_D &=& \fr{2 E_0^2 }{\pi e u} \lambar^2 \sin^2(2\pi
K)\Gamma^2(1-2K)
  \left|\fr{ \omega^2 -
    Q^2}{E_0^2} \right|^{2 K -1} \times\\
& & \mbox{sign}(\omega)\:,
\eea
whereas the drag becomes exponentially suppressed for $|Q| - |\omega|
\gg T$.
This result has been already derived by Nazarov and Averin
\cite{nazarov_averin}.
The seeming $|\omega^2-Q^2|^{2K-1}$
singularity it exhibits near the threshold
$|\omega| = |Q|$ is smeared over a regime where $||\omega| -
|Q|| \lesssim T$, which can be quantitatively described by Eq.
\Ref{drag_field}.
For large currents, $|\omega| \gg |Q|$, the drag goes as $\eps_D
\propto
\omega^{4K-2}$. Fig. \ref{fig-drag} illustrates this
behaviour.

\subsection{Linear drag resistance}\label{subsec-linear_drag}
\begin{figure}
  \begin{center}
    \epsfxsize=6.0cm
    \leavevmode
    \epsffile{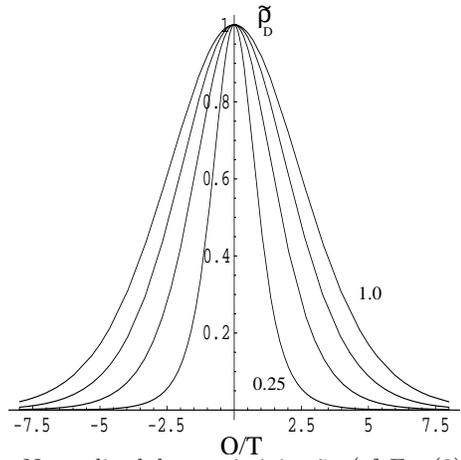}
    \caption{
      Normalized drag resistivity $\tilde \rho_D$
(cf.~Eq.~\Ref{norm_drag}) as
      function of $Q/T = u \delta k_F /T $ for interaction parameter
      $K=1$, 0.75, 0.5 and 0.25.
respectively.
      }
    \label{fig-peak}
  \end{center}
  \vspace{-0.0cm}
\end{figure}

{From} Eq. \Ref{drag_field} one easily finds the linear drag
resistivity
$\rho_D  = \fr{d \eps_D }{ dI}|_{I=0}$ as
\ben\label{drag_resistance}
\rho_D =  \rho_0  \left( \fr{ T}{E_0} \right)^{4K - 3} \left(
  \Nprime{\fr{Q}{T}}{K} \N{-\fr{Q}{T}}{K} + \N{\fr{Q}{T}}{K}
  \Nprime{-\fr{Q}{T}}{K}
\right)\:,
\een
with $\rho_0 =  \sqrt{8} \pi^{4K-3} E_0 \lambar^2 / e^2 u$
and $\Nprime{r}{K} \equiv d \N{r}{K} /dr$.
The drag resistivity is a symmetric function of
$Q$ with its maximum at $Q = 0$, where
$\rho_D \propto T^{4K-3}$.
For $|Q|$ larger than temperature the linear drag resistance is
exponentially suppressed, due to the exponential suppression of
$\N{r}{K}$ for large and positive $r$.
Thus, the peak of $\rho_D(Q)$ has
height $\propto T^{4K-3}$ and its width is directly proportional
to temperature.
More details can be extracted from Fig. \ref{fig-peak}, where
the normalized drag resistance
\ben\label{norm_drag}
\tilde \rho_D(Q/T) \equiv \fr{\rho_D(Q,T) }{\rho_D(Q=0,T) }
\een
is plotted against $Q/T$ for $K=1$, 0.75, 0.5 and 0.25.

\subsubsection{Influence of electron spin}
A spin-full system can perturbatively be treated in the same way as
above the spin-less system. The resulting expressions for drag
$\eps_D$ and drag resistance $\rho_D$ are basically identical to
\Ref{drag_field} and \Ref{drag_resistance}, except of one difference,
which is essential: The interaction-determined constant $K$ is replaced
by an effective
$\tilde K = \fr{1}{2}(K_s+ K_{c-})$, where $K_s$ is the
interaction parameter of the spin modes (cf. \cite{klesse_stern}).
For weak backscattering coupling $K_s \approx 1$, such that
$\tilde K = \fr{1}{2}(1 + K)$. I.e. the effective parameter $\tilde K$
of the spin-full system is closer to the non-interacting
(Fermi-liquid) value $K_{Fl} = 1$, which 
reflects the moderating effect of the spin-modes.

\section{Incommensurate-Commensurate
transition}\label{sec-ic-transition}

Eq. \Ref{grand_canonical} is the Hamiltonian  of the sine-Gordon model
with an additional coupling of the soliton density $ n_s  = \int dx
\: \dx \sqrt{8}  \phi_{c-} / 2 \pi L$ to the potential difference
$W$. It is this coupling which gives rise to the IC transition
\cite{pokrovsky,schulz}:

At large $\bar \lambda E_0 $ and small $|W|$ the $\cos$-term
dominates and suppresses the soliton density $n_s$ to zero, which
is the commensurate phase. In the opposite situation, the $\cos$-term
is negligible and the ground state density of the solitons
is adjusted to $n_s = \pi K W/ u$.
This value of $n_s$ corresponds to electron densities
$n_w \equiv \av{\varrho_w}$ in the two wires that are independently
tuned
by the respective electro-statical potentials $V_w$.
This is the incommensurate phase.

In the classical model, for $W$ less than the energy $E_s$ of a single
soliton the density $n_s$ vanishes strictly at zero temperature.
The transition to the incommensurate phase sets in
at $|W| = E_s$, where for small $|W|-E_s > 0$ the soliton density
increases logarithmically with $n_s \sim |\ln(|W|-E_s)|\inv$,
corresponding to a critical exponent $\beta = 0$
\cite{pokrovsky,schulz}.
At finite temperatures the transition is smeared out, as ususal
for 1D statistical models.

Quantizing the model modifies the transition in two aspects: the
transition energy increases to a renormalized soliton energy $\bar
E_s = \Delta$ and the critical exponent changes to $\bar \beta =
1/2$ \cite{pokrovsky,schulz}.

Obviously, the IC transition is accompanied by a drastic change in
the inter-wire electronic correlations. It can be therefore
expected that the Coulomb drag varies strongly near the
transition, and we now address this variation.

In the commensurate phase the electron densities in the two wires
are inter-locked to one another. The commensurate phase as a
gapped phase has no low-energetic excitations and therefore shows
no response to a small anti-symmetric field $\eps_{c-}$ at
strictly vanishing temperature. Hence, at linear response the
resistivity to the flow of non-identical currents in the two wires
is infinite, and so is the drag resistivity.

At finite but small temperatures $T \ll \Delta - |W|$ solitons and
antisolitons are thermally activated with a density $n_{s/a}
\propto \exp(-E_{s/a}/T)$, where $E_{s/a} = \Delta \mp W$ denotes
the energy of soliton and antisoliton in the external potential.
 In the presence
of an anti-symmetric field $\eps_{c-}=\eps_1=-\eps_2$ both
solitons and antisolitons contribute to an anti-symmetric current
$j_{c-}$. A detailed calculation of this current response to the
field $\eps_{c-}$, taking into account the damping of the soliton
motion is beyond our present scope. However, we are able to
conjecture that the temperature dependence of the anti-symmetric
conductance is exponentially activated, following the temperature
dependence of the soliton density. The conductance is the product
of the soliton density by their mobility. As long as the soliton
density is small, we expect the mobility to be limited by the
interaction of each soliton with its environment, rather than by
the interaction with other solitons. In that limit the mobility
will not depend on the soliton density, and thus will not be
thermally activated. In that limit, then, we expect an
exponentially large drag resistivity \ben\label{exp_drag}
 \rho_D \sim a e^{(\Delta-|W|)/T}
\een at low temperatures $T \ll \Delta - |W|$, where the double
wire system is in the commensurate phase. The coefficient $a$ may
still exhibit a significant
temperature dependence.

When the external potential $|W|$ exceeds the critical value
$\Delta$ for the incommensurabiltity transition, proliferation of
non-thermal solitons ($W>\Delta$) or antisolitons ($W < -\Delta$)
sets in with a density increasing as $n_{s/a} \propto
(W^2-\Delta^2)^{1/2}$ \cite{pokrovsky,schulz}. Accordingly, above
the transition point the conductivity $\sigma_{c-}$ increases
$\propto (W^2-\Delta^2)^{1/2}$. This means that the exponential
behaviour of the drag resistivity \Ref{exp_drag} changes to an
inverse square root dependency \ben\label{square_root} \rho_D \sim
b (W^2 - \Delta^2)^{-1/2}\:, \een valid at temperatures $T \ll |W|
- \Delta $. Also the coefficient $b$ may exhibit a significant
temperature dependence. The transition from the exponential
behaviour \Ref{exp_drag} to the inverse square root dependence
takes place as $W$ passes through an energy window of width of the
order of $T$ at $\pm \Delta$.

\begin{figure}
  \begin{center}
    \epsfxsize=7.0cm
    \leavevmode
    \epsffile{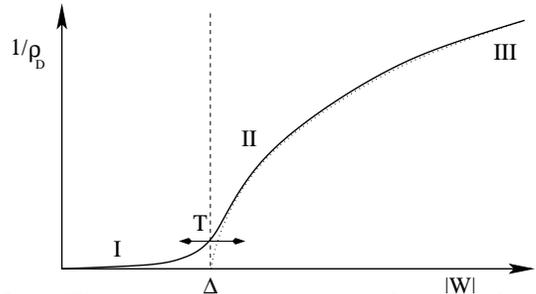}
    \caption{ The inverse drag resistivity of a pair of wires at low
    temperature, $T \ll \Delta $, as a function of
    the potential difference $W$ (schematic).
    For $|W| $ less than the energy gap $\Delta$ the charges order in
a
    zigzag formation (commensurate phase) and the drag is
exponentially
    large, $\rho_D \sim a \exp(\Delta-|W|)/T$ (I). For $W$ above the
gap,
    $|W|-\Delta \gg T$, the drag resistance decreases as $ (|W| -
    \Delta)^{-1/2}$ with increasing $W$ (II). At large $W \gg \Delta$
the
    drag resistance decays linearly with $W$ (III).}
    \label{fig1}
  \end{center}
  \vspace{-0.5cm}
\end{figure}

At large external potentials, $|W| \gg \Delta$, the (anti)soltion
density is nearly linear in $|W|$, such that, assuming that the
soliton mobility is still independent of its density, the drag
resistance goes as $\rho_D \propto 1/|W|$.
 The low-temperature behaviour of the drag resistivity is summarized
in
Fig. \ref{fig1}.

\subsubsection{Influence of electron spin}
So far the discussion ignores electron spin and is therefore
limited to a spin-polarized system. What will happen for unpolarized
electrons?
Referring to previous work \cite{klesse_stern} the following
can be said: The spin fluctuations moderate correlations of the charge
modes, such that $\Delta$ assumes a much lower value than in a
spin-polarized system. If $T\ll \Delta$  can nevertheless be realized,
we
expect qualitatively the same characteristics of the drag as in the
spin-less
case.
This can be seen by the renormalization group equations of a spin-full
system, which suggest that both spin and relative charge modes become
gapped at low temperatures (cf. \cite{klesse_stern}). Inspection of
the
backscattering hamiltonian (see Eqs. (41-43) in \cite{klesse_stern})
reveals that also in this case solitonic excitations in the $c-$ mode
are
possible. This leads to the same phenomenology of the low temperature
drag as just discussed for the spin-less case.

\section{Discussion}\label{discussion}
We have seen that the incommensurate-commensurate transition of
a strongly coupled double wire is accompanied by a marked change of
the drag resistance (cf.~Eq.s~\Ref{exp_drag} and
\Ref{square_root}). Observation of this effect requires that
the temperature is of order of or below the energy gap $\Delta$.
Taking the estimate of $\Delta$ of order $mK$ for a spin-unpolarized
double
wire with $E_0 \approx 10K$ \cite{klesse_stern} we conclude that the
strongly coupled regime might be elusive for these systems. However,
in case of spin-polarized
electrons the situation looks much better, where for the same
parameters one obtains $\Delta$ of order $100mK$.

The perturbative results derived in Sec. \ref{sec-high-temperature}
should be applicable to existing experimental data
\cite{debray,debray_review,yamamoto}. In fact, both groups observe
a peak of the
drag resistivity as the gate voltages are varied. While the
qualitative
agreement is obvious, a quantitative comparison is difficult because
the relation between electron densities in the wires and
gate voltages is not trivial.
Eq. \Ref{drag_field} suggests that measurement of the
non-linear drag might be an insightful probe in future experiments.

It is a pleasure to acknowledge helpful and constructive conversations
with P. Debray, L. Glazman, W. H\"ausler, and A. Rosch. We thank the
Minerva-Foundation (Munich) and the Sonderforschungsbereich/TR 12 for
financial support.


\begin{thebibliography}{10}
\bibitem{units} Throughout the paper we use units in which $\hbar =1$
    and $k_B =1$.
\end{thebibliography}
\end{document}